\documentclass[11pt,letterpaper]{article}   %% LaTeX 2e (preferred)
\usepackage{osajnl} %% do not use with REVTeX4
\usepackage{setspace}

\begin{document}

\title{The one-dimensional holographic display}

\author{Kim Young-Cheol}
%% for REVTeX4, each author name can be set in a separate \author{} field

%\address{Optical Society of America, 2010 Massachusetts Avenue, NW, \\
%Washington, D.C. 20036} %\affiliation will also work

%\email{kyc@linepicture.com}

\begin{abstract}This paper introduces a new concept of one-dimensional
hologram which represents one line image, and a new kind of display structure using it.
This one-dimensional hologram is similar to a superpositioned diffraction lattice.
And the interference patterns can be efficiently computed with a simple optical
computing structure. This is a Proposal for a new kind of display method.
\end{abstract}
% REPLACE WITH CORRECT OCIS CODES FOR YOUR ARTICLE
                          % NOTE: \ocis{} IS ALIASED TO \pacs{} BUT MUST
                          % FORMAT THE TERMS CORRECTLY FOR EACH JOURNAL

%\maketitle %% null function with osajnl.sty

\section{Introduction}
 This paper intends to introduce a new
 display method using the holography theory by Dennis Gabor in 1948.
 The holography had been expected to be a popular display method
 for a 3-dimensional image, but the burden of tremendous amount of data processing
 prohibited the practical application. Thus, I would like to introduce
 a one-dimensional holographic display concept which can reduce the
 burden of data processing, can adopt simple optical computing method, and
 has some more practical merits in manufacturing. A one-dimensional hologram
 can display only a two-dimensional image, but it does not require
 the lenses like HMD(Head Mount Display). Instead,
 this one-dimensional holographic display device has a possibility of
 showing a real-time two-dimensional information without a lens,
 within today's technology.

 This paper contains theoretical considerations about the one-dimensional
 Holography, and the equations that I derived showing the existence of
 the one-dimensional hologram as well as discussions about practical
 structures of the display device, the light modulators, and the optical computing device.

 This work had started by considering the information dimension of a
 hologram. A traditional two-dimensional hologram can display a three
 dimensional image, So I speculated that this 2 to 3 relationship
 between data dimension and image dimension could be transformed into 1 to
 2 relationship. So, I first tried to find a
 one-dimensional hologram for a two-dimensional image by geometrical
 method, but failed. Instead, I found that a diffraction lattice like
 one-dimensional hologram is formed by some special condition of line
 image. And, I had conceived a vector and matrix based mathematical
 technique which can easily express the idea. Unexpectedly, this
 technique was also useful to handle the problem of diffraction
 efficiency and noise cancelling problem for computed
 artificial two-dimensional hologram.
 
\section{The Hologram}
A phasor expression for a wave from one point source is Eq. (1).

Let $A(\vec{r})$ represent the relative phase and the amplitude of a wave function, then
\begin{equation} %equation 1
A( \vec{r} )= \alpha ( \vec{r} ) \exp{(i[f(\vec{r}) + \delta])} 
\end{equation}
($ \alpha( \vec{r} )=\sqrt{\mbox{the detection probability of wave
 quanta}} $ , $ f( \vec{r} )=2 \pi r / \lambda $  )
 The $\alpha$ is indeed a complex function, but when $ r \gg \lambda
  $, it is a simple function proportional to $1/r$, and
it is actually a constant when computing a one-dimensional hologram mainly discussed in this paper.
The traditional wave function $\Psi$ is obtained by considering the time term.
$ \Psi ( \vec{r},t ) = A( \vec{r} )\exp(2 \pi i \omega t)$
The above phasor expression does not represent a real wave, but the
 interference pattern of a certain point depends on only relative phase differences
 between light rays, so the time term disappears when computing
 the interference pattern. The coherent rays have constant relative
 phases. The polarization of lights are ignored.
A phasor expression for waves from many point sources is Eq. (2)
 by the principle of superposition.

Let $S(\vec{r})$ be superpositioned $A(\vec{r})$ of Eq.(1), then
\begin{equation} %equation 2
S( \vec{r} )= \sum \alpha ( \vec{r} ) \exp{(i[f(\vec{r}) + \delta])}
\end{equation}
An actual hologram is a record of the interference pattern on a photographic
plate. The interference pattern depends on the illumination.

Let $I(\vec{r})$ be the illumination over the space then,
\begin{equation} %equation 3
I(\vec{r})=\left| S(\vec{r}) \right|^2= \left| \sum \alpha (\vec{r})
\exp{(i[f(\vec{r}) + \delta])} \right|^2 
\end{equation}
This can be rewritten as
\begin{equation} %equation 4
I(\vec{r})=  \sum \alpha (\vec{r}) \exp{(i[f(\vec{r}) + \delta])}
\times \sum \alpha (\vec{r}) \exp{(-i[f(\vec{r}) + \delta])}
\end{equation}
And, when expanded to a matrix, it is
%equation 5
\begin{equation} I(\vec{r})=
\left(\begin{array}{ccccc}
\alpha_{1}^2 & \alpha_{1}\alpha_{2}e^{i(f_1-f_2)} & \alpha_{1}\alpha_{3}e^{i(f_1-f_3)} & \alpha_{1}\alpha_{4}e^{i(f_1-f_4)} & \cdots \\
\alpha_{2}\alpha_{1}e^{i(f_2-f_1)} & \alpha_{2}^2 & \alpha_{2}\alpha_{3}e^{i(f_2-f_3)} & \alpha_{2}\alpha_{4}e^{i(f_2-f_4)} & \cdots \\
\alpha_{3}\alpha_{1}e^{i(f_3-f_1)} & \alpha_{3}\alpha_{2}e^{i(f_3-f_2)} & \alpha_{3}^2 & \alpha_{3}\alpha_{4}e^{i(f_3-f_4)} & \cdots \\
\alpha_{4}\alpha_{1}e^{i(f_4-f_1)} & \alpha_{4}\alpha_{2}e^{i(f_4-f_2)} & \alpha_{4}\alpha_{3}e^{i(f_4-f_3)} & \alpha_{4}^2 & \cdots \\
\cdots & \cdots & \cdots & \cdots & \cdots
\end{array} \right)
\end{equation}
 This matrix needs normalization for actual application, but the image reproduction with a
 hologram may now be certified. If you select $\alpha_1\exp(if_1(\vec{r}))$ as reference light
 and illuminate it as reproducing light(select $\alpha_1=1$ for
 intensity) on a hologram which represents the above matrix, then
 $J(\vec{r})$ represents the modulated lights,
%equation 6
\begin{eqnarray}
\lefteqn{J(\vec{r})} \nonumber \\
& = & I(\vec{r})\exp{(if_1(\vec{r}))} \nonumber \\
& = &
\left(
\begin{array}{ccccc}
e^{if_1} &
\alpha_2e^{i(2f_1-f_2)} &
\alpha_3e^{i(2f_1-f_3)} &
\alpha_4e^{i(2f_1-f_4)} &
\cdots \\
\alpha_2e^{if_2} &
\alpha_2^2e^{if_1} &
\alpha_{3}\alpha_{3}e^{i(f_1-(f_3-f_2))} &
\alpha_{2}\alpha_{4}e^{i(f_1-(f_4-f_2))} &
\cdots \\
\alpha_3e^{if_3} &
\alpha_{3}\alpha_{2}e^{i(f_1+(f_3-f_2))} &
\alpha_3^2e^{if_1} &
\alpha_{3}\alpha_{4}e^{i(f_1-(f_4-f_3))} &
\cdots \\
\alpha_4e^{if_4} &
\alpha_{4}\alpha_{2}e^{i(f_1+(f_4-f_2))} &
\alpha_{4}\alpha_{3}e^{i(f_1+(f_4-f_3))} &
\alpha_4^2e^{if_1} &
\cdots \\
\cdots & \cdots & \cdots & \cdots & \cdots
\end{array}  \right)\nonumber\\
\end{eqnarray}
 Take this matrix's diagonals to the first term, remaining first
column to the second term, remaining first row to the third term, and others
are added to their symmetry conjugated and defined by cosine function,
then the result is
\begin{eqnarray} %equation 7
J(\vec{r}) & = & \exp{(if_1(\vec{r}))}(1+\sum\alpha^2) \nonumber\\
& & + \sum \alpha \exp{(if(\vec{r}))} \nonumber\\
& & + \sum \alpha \exp{(i[2f_1(\vec{r})-f(\vec{r})])}\nonumber\\
& & + 2\exp{(if_1(\vec{r}))} \sum \alpha_{m}\alpha_{n} \cos(f_m(\vec{r})-f_n(\vec{r}))
\end{eqnarray} 
The first term is 0th order term,  the second term represents image,
the third term represents conjugate image, and the fourth term is noise term.
The above method does not depend on any particular coordinate system,
so it can explain volume hologram, as well. 

\section{The One-dimensional Hologram}
When considering the dimensions of storing and displaying hologram,
the volume hologram can display three-dimensional image, and can be
multiplexed with wavelengths and spatial coordinates of light sources.
The two-dimensional hologram can display three-dimensional image,
and can not be multiplexed.
When considering one-dimensional hologram, one-dimensional hologram
can display one-dimensional image(one-line image), and can not be
multiplexed.

In Cartesian coordinate system, when $r\gg\lambda$, Eq. (1) can be
rewritten as
\begin{eqnarray} %equation 8
A(\vec{r}) & = & A(X,Y,Z) \nonumber \\
& = & \frac{\alpha}{r}\exp{( 2\pi i[{\frac{\sqrt{(Ox-X)^2+(Oy-Y)^2+(Oz-Z)^2}}{\lambda}}+\delta ])}  
\end{eqnarray}
($\sqrt{(Ox-X)^2+(Oy-Y)^2+(Oz-Z)^2}=r$, $(Ox, Oy, Oz)$ is
 the coordinate of dot light source) This expression represents a volume hologram's
 case. The two-dimensional hologram formation on a $XY$ plain can be obtained
 by substitution $Z=0$. The thickness of zero can not exist in real world,
 so an actual two-dimensional hologram by photographic method
 is a thin volume hologram and in fact, it is significantly
 advantageous to reduce image noise.

 If $Y$ is substituted with zero again, it could be called as a one-dimensional hologram.
 But, it is a hologram on a physical line. It is hard to find physical
 meaning. Instead, if a hologram on a plain is expressed with single
 axes information, then it also can be called as the one-dimensional hologram.

The phase of Eq. (8) is relative to the source of light. It is
possible to transform the expression to be relative to the origin
$(X,Y,Z)=(0,0,0)$ of coordinate system.
When the light is parallel, the $r$ of Eq. (1) is infinite, $r_0$ is the
distance between the origin and source.
\begin{eqnarray*}
A(\vec{r}) & = & A(X,Y,Z) \\
& = & \frac{\alpha}{r}\exp{(2 \pi i [\frac{ r(x,y,z)-r_0}{\lambda} +\delta ])} \\
& = & \frac{\alpha}{r}e^{\textstyle 2 \pi
  i[\frac{\sqrt{(Ox-X)^2+(Oy-Y)^2+(Oz-Z)^2}- \sqrt{Ox^2+Oy^2+Oz^2}} {\lambda} +\delta ]} \\
& \approx & \frac{\alpha}{r}\exp{( 2 \pi i[\frac{
    -\frac{Ox X + Oy Y + Oz Z}{\sqrt{Ox^2+Oy^2+Oz^2}}} {\lambda} +\delta ])}\\
& = & \frac{\alpha}{r}\exp{( -2 \pi i[\frac{
    \frac{\textstyle\vec{r}\cdot\hat{x}X + \vec{r}\cdot\hat{y}Y + \vec{r}\cdot\hat{z}Z}{\textstyle r}}{\lambda} +\delta ])} \\
& = & \frac{\alpha}{r}\exp{( -2 \pi i[\frac{\textstyle\hat{r}\cdot\hat{x}X + \hat{r}\cdot\hat{y}Y + \hat{r}\cdot\hat{z}Z}{\lambda} +\delta ])}
\end{eqnarray*}
 When $r=\infty$, $\alpha/r$  can be changed to constant $\alpha$, therefore
\[ A(X,Y,Z)=\alpha \exp{( -2 \pi i[\frac{\hat{r}\cdot\hat{x}X + \hat{r}\cdot\hat{y}Y + \hat{r}\cdot\hat{z}Z}{\lambda} +\delta ])}\]
 Now, one dimension can be reduced by limiting $XY$ plain with
 $Z=0$. Therefore the result is,
\begin{equation} %equation 9
A(X,Y)=\alpha \exp{( -2 \pi i[\frac{\hat{r}\cdot\hat{x}X + \hat{r}\cdot\hat{y}Y}{\lambda} +\delta ])}
\end{equation}
($\alpha \exp{( -2 \pi
  i[\frac{\textstyle\hat{r}\cdot\vec{P}}{\lambda} +\delta ])} $ is more
adept for final result, but Eq. (9) shall be used for convenience)
 According to the method of Eq. (2), the expression for many points is
\[S(X,Y)=\sum \alpha \exp{( -2 \pi i[\frac{\textstyle\hat{r}\cdot\hat{x}X + \hat{r}\cdot\hat{y}Y}{\lambda} +\delta ])}\]
  At this time, when $\hat{r}\cdot\hat{y}=c$(onstant) (all the points
  are on same latitude in polar coordinate), the above can be rewritten as
\begin{equation} %equation 10
S(X,Y)=\exp{( -2 \pi i\frac{ \hat{r}\cdot\hat{y}}{\lambda}Y)}
\sum \alpha \exp{( -2 \pi i[\frac{\hat{r}\cdot\hat{x}}{\lambda}X +\delta ])}
\end{equation}
The real hologram information is obtained by applying the method of Eq. (4) for
Eq. (10). Let $\alpha_1=1$, then the result is,
\begin{eqnarray} %equation 11
\lefteqn{I(X,Y)} \nonumber \\
& = & |S(X,Y)|^2 \nonumber \\
& = & \exp{( -2 \pi i
\frac{\hat{r}\cdot\hat{y}}{\lambda} Y )}
\sum \alpha \exp{( -2 \pi  i[\frac{\hat{r}\cdot\hat{x}}{\lambda}X
  +\delta ])} \nonumber\\
& & \times  \exp{( 2 \pi i\frac{
    \hat{r}\cdot\hat{y}}{\lambda}Y)}\sum \alpha \exp{( 2 \pi
  i[\frac{\textstyle\hat{r}\cdot\hat{x}}{\lambda}X +\delta ])}\nonumber
\\
& = & \sum \alpha \exp{( -2 \pi
  i[\frac{\hat{r}\cdot\hat{x}}{\lambda}X +\delta ])}\times
\sum \alpha \exp{( 2 \pi
  i[\frac{\textstyle\hat{r}\cdot\hat{x}}{\lambda}X +\delta ])} \nonumber \\
& = & I(X) \nonumber \\
& = & \left(
\begin{array}{cccc}
1 & 
\alpha_2e^{2 \pi i
\frac{\textstyle (\hat{r_1}-\hat{r_2} )\cdot\hat{x}}{\textstyle \lambda}X} &
\alpha_3e^{2 \pi i
\frac{\textstyle  (\hat{r_1}-\hat{r_3} )\cdot\hat{x}}{\textstyle \lambda}X} &
\cdot \\
\alpha_2e^{2 \pi i
\frac{\textstyle  (\hat{r_2}-\hat{r_1} )\cdot\hat{x}}{\textstyle \lambda}X} &
\alpha_2^2 &
\alpha_{2}\alpha_{3}e^{2 \pi i
\frac{\textstyle  (\hat{r_2}-\hat{r_3} )\cdot\hat{x}}{\textstyle \lambda}X} &
\cdot \\
\alpha_3e^{2 \pi i
\frac{\textstyle  (\hat{r_3}-\hat{r_1} )\cdot\hat{x}}{\textstyle \lambda}X} &
\alpha_{3}\alpha_{2}e^{2 \pi i
\frac{\textstyle \! (\hat{r_3}\!\!-\hat{r_2}\! )\!\cdot\hat{x}}{\textstyle  \lambda}X} &
\alpha_3^2 &
\cdot \\
\cdots & \cdots & \cdots & \cdot
\end{array} \right) \nonumber \\
\end{eqnarray}
The term of $Y$ was cancelled. So, this is expressed with
one-dimensional data which depend on $x$ axes only.
Therefore, Eq. (11) represents a one-dimensional hologram in this paper.
According to the method of Eq. (5),(6), the modulation of reproducing
light is
\begin{eqnarray*}
\lefteqn{J(X,Y)}\\
& = & \exp{( -2 \pi i\frac{\hat{r}_1\cdot\hat{x}X +
    \hat{r}_1\cdot\hat{y}Y}{\lambda})} \\
&   &
\left(\begin{array}{cccc}
1 & 
\alpha_2e^{2 \pi i
\frac{\textstyle (\hat{r_1}-\hat{r_2} )\cdot\hat{x}}{\textstyle \lambda}X} &
\alpha_3e^{2 \pi i
\frac{\textstyle  (\hat{r_1}-\hat{r_3} )\cdot\hat{x}}{\textstyle \lambda}X} &
\cdot \\
\alpha_2e^{2 \pi i
\frac{\textstyle  (\hat{r_2}-\hat{r_1} )\cdot\hat{x}}{\textstyle \lambda}X} &
\alpha_2^2 &
\alpha_{2}\alpha_{3}e^{2 \pi i
\frac{\textstyle  (\hat{r_2}-\hat{r_3} )\cdot\hat{x}}{\textstyle \lambda}X} &
\cdot \\
\alpha_3e^{2 \pi i
\frac{\textstyle  (\hat{r_3}-\hat{r_1} )\cdot\hat{x}}{\textstyle \lambda}X} &
\alpha_{3}\alpha_{2}e^{2 \pi i
\frac{\textstyle \! (\hat{r_3}\!\!-\hat{r_2}\! )\!\cdot\hat{x}}{\textstyle  \lambda}X} &
\alpha_3^2 &
\cdot \\
\cdots & \cdots & \cdots & \cdot
\end{array} \right) \nonumber \\
\end{eqnarray*}
And, sorting as Eq. (7), results are
\begin{eqnarray} %equation 12
\lefteqn{J(X,Y)}\nonumber\\
&=& \exp{( -2 \pi i\frac{\hat{r}_1\cdot\hat{x}X +
    \hat{r}_1\cdot\hat{y}Y}{\lambda})}(1+\sum\alpha^2)
\mbox{......................................\{1\}} \nonumber \\
& & + \sum \alpha \exp{( -2 \pi i\frac{\hat{r}\cdot\hat{x}X +
    \hat{r}_1\cdot\hat{y}Y}{\lambda})}
\mbox{..............................................\{2\}} \nonumber \\
& & + \sum \alpha \exp{( -2 \pi
  i\frac{(2\hat{r}_1-\hat{r})\cdot\hat{x}X +
    \hat{r}_1\cdot\hat{y}Y}{\lambda})}
\mbox{..................................\{3\}} \nonumber \\
& & + 2\exp{( -2 \pi i\frac{\hat{r}_1\cdot\hat{x}X +
    \hat{r}_1\cdot\hat{y}Y}{\lambda})} \sum \alpha_{mn} \cos{(2 \pi
\frac{(\hat{r}_m-\hat{r}_n)\cdot\hat{x}X}{\lambda})}
\mbox{...\{4\}}\nonumber\\
\end{eqnarray}
($\delta$ is omitted) Also,
\{1\} is term of 0th order, \{2\} is a term representing the image,
\{3\} is a term representing the conjugated image,
which confirms that it works as a hologram.
Some of the lights expressed by terms of \{3\} and \{4\}, may not be
reproduced because the final unit vectors of light ray always have to
satisfy the size of 1. This means, for example, among the lights of term
\{3\}, the lights those $((2\hat{r}_1-\hat{r})\cdot\hat{x})^2 +(\hat{r}_1\cdot\hat{y})^2$
are smaller than 1, can be generated.

It is the same situation of a diffraction grating that is expressed with
grating equation$^1$ $ \alpha (\sin{\theta_m}-\sin{\theta_i}) = m\lambda$.
In grating equation, the degree of $\pm m$ is limited as the absolute
value of a sinusoidal function is limited to 1.

Term \{2\} can have physical meaning with different wavelengths or
latitude angles of incidence, so it is impossible to multiplex the
one-dimensional hologram by wavelengths or latitude angles of incidence.

\section{The one-dimensional holographic display device}
The one-dimensional hologram may be used to make a display device as
described in figure 1.
  \begin{figure}[h]%figure1
  \centering
  \includegraphics[angle=90, width=5in, height=3in]{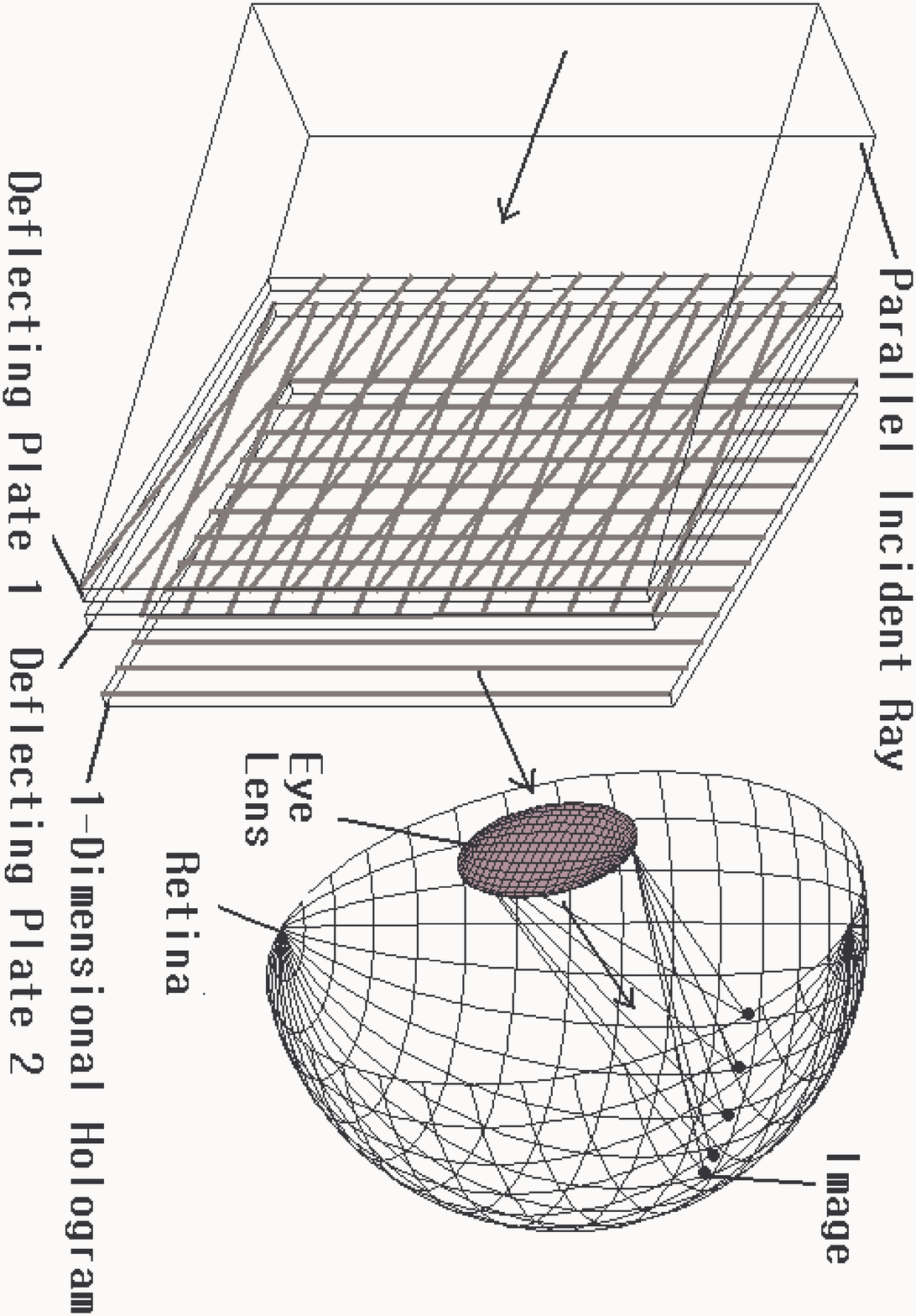}
  \caption{An example of one-dimensional hologram display device.}
  \end{figure}
To reproduce a image, a one-dimensional hologram should be expressed
with a spatial light modulator and a proper reproducing light should
be illuminated, then one line of image shall be displayed. And, the
whole plain image is displayed by updating the one-dimensional
hologram and the angle of incidence( $ \hat{r}_1 \cdot \hat{y} $ of
Eq. (12) term \{2\}) of the parallel reproducing light synchronously
and in sequence. The natural color is obtained by repeating display with
the three primary colors.

The incident angle of reproducing light should be adjustable, so a
deflection device is needed. There may be many kind of deflection
device, but the one-dimensional hologram itself also can be used as a
deflector. In fact, the one-dimensional hologram deflector is
identical to a cosine diffraction lattice.

The deflecting plate 1, 2 and the one-dimensional hologram are cross
structured light modulators. One of the deflecting plate 1 or 2
operates at a time and the other maintains the transparent state.
The incident angle of parallel ray in figure 1 is fixed as in
figure 2, then one of the deflectors 1 and 2
deflects the parallel ray by deflection range 1 or 2 in figure 2.
  \begin{figure}[h]%figure2
  %\begin{center}
  \centering
  \includegraphics[angle=90, width=5in, height=3.25in]{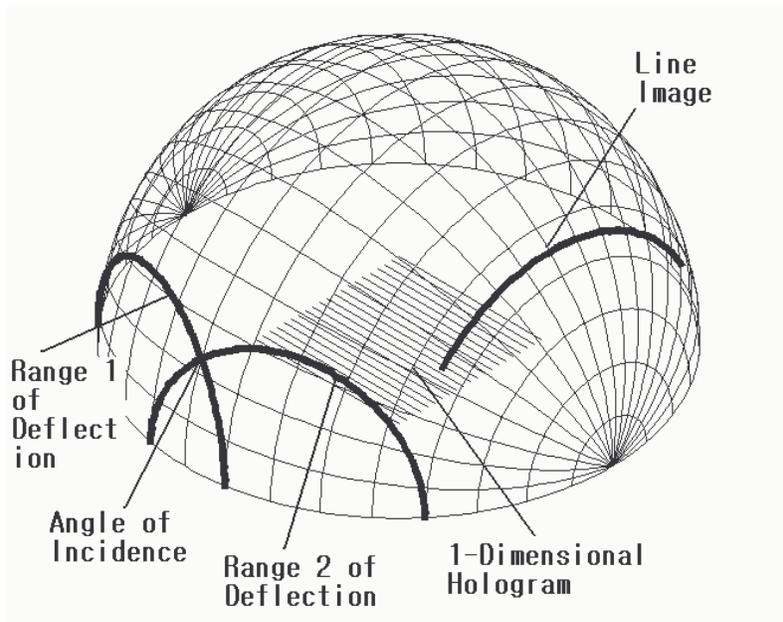}
  \caption{Deflection ranges of deflecting plates, and displaying image.}
  \end{figure}
This structure makes it possible to eliminate the 0th order light by
total internal reflection. It is considerable to replace one of the
one-dimensional hologram deflectors with a multiplexed volume hologram.

Theoretically, it is possible to display a two-dimensional image with
above scheme, but there are some more considerable problems to
actually develop and operate this display device. They are developing
optical modulation device, noise cancelling of image,
and the fast computation of the interference pattern.
And, the comparison with conventional
two-dimensional hologram method or with controllable diffraction lattice
method is needed to verify the usefulness of the one-dimensional
hologram display method.

\subsection{The light modulating device}
A hologram display device requires very high resolution spatial light
modulator than conventional display device. Recently, it had been announced that
liquid crystal display device has reached the resolution of
$10\mu m$. However, this resolution is still not enough to display a
hologram.

The hologram method display device has no relation
between the image resolution and the resolution of optical modulation
device. The resolution of modulator is related to the field of view,
precisely, it is related to the angle between a light from a
picture element of an image and the reference light of
hologram.  When $\alpha_1=1$, Eq. (11) can be rewritten as,
\begin{eqnarray}%equation13
\lefteqn{I(X)} \nonumber \\
& = & \left(
\begin{array}{cccc}
1 & 
\alpha_2e^{2 \pi i
\frac{\textstyle (\hat{r_1}-\hat{r_2} )\cdot\hat{x}}{\textstyle \lambda}X} &
\alpha_3e^{2 \pi i
\frac{\textstyle  (\hat{r_1}-\hat{r_3} )\cdot\hat{x}}{\textstyle \lambda}X} &
\cdot \\
\alpha_2e^{2 \pi i
\frac{\textstyle  (\hat{r_2}-\hat{r_1} )\cdot\hat{x}}{\textstyle \lambda}X} &
\alpha_2^2 &
\alpha_{2}\alpha_{3}e^{2 \pi i
\frac{\textstyle  (\hat{r_2}-\hat{r_3} )\cdot\hat{x}}{\textstyle \lambda}X} &
\cdot \\
\alpha_3e^{2 \pi i
\frac{\textstyle  (\hat{r_3}-\hat{r_1} )\cdot\hat{x}}{\textstyle \lambda}X} &
\alpha_{3}\alpha_{2}e^{2 \pi i
\frac{\textstyle \! (\hat{r_3}\!\!-\hat{r_2}\! )\!\cdot\hat{x}}{\textstyle  \lambda}X} &
\alpha_3^2 &
\cdot \\
\cdots & \cdots & \cdots & \cdot
\end{array} \right) \nonumber \\
& = & 2 \sum \sum \alpha_{m}\alpha_{n} \cos{(2
  \pi\frac{\textstyle(\hat{r}_m-\hat{r}_n)\cdot\hat{x}X}{\lambda})}
\end{eqnarray}

This shows that a hologram is the sum of spatial periodic structures
which is expressed with $\lambda/(\hat{r}_m-\hat{r}_n)\cdot\hat{x}$.
The possible maximum value of $(\hat{r}_m-\hat{r}_n)\cdot\hat{x}$ is 2
and at least two pixel is needed to express one spatial period, so,
the resolution of light modulator should be $\lambda/4$ to display a
image without the limitation of visual field. To express natural color,
if about $400nm$ of blue ray wavelength is substituted for $\lambda$, then
a modulator of approximately $100nm$ resolution is required.
When using previously mentioned liquid crystal display device of
$10\mu m$ resolution as spatial light modulator, from
$\frac{\textstyle 400nm}{\textstyle 2(\hat{r}_m-\hat{r}_n)\cdot\hat{x}}=10\mu m$,
the maximum field of view is
$(\hat{r}_m-\hat{r}_n)\cdot \hat{x}=\frac{\textstyle 400nm}{\textstyle
  2 \times 10 \mu m}=0.02$, this is capable of displaying about $2cm$
wide virtual screen at $1m$ distance, which has no practical use.
Fortunately, there have been continuous researches for other types of
optical modulation devices. As one of them, according to recently
opened Japan NTT Docomo's patent document$^2$, they
have mentioned that higher than $1\mu m$ resolution may be obtained by
using a photo-refractive crystal. This is capable of displaying about
$20cm$ wide screen at $1m$ distance, but still it is not fully enough.

The $1\mu m$ resolution mentioned above is the possible resolution for
the two-dimensional hologram.
The resolution of modulator can be improved by using one-dimensional hologram.
To display a two-dimensional hologram, one pixel electrodes should be placed for
each pixel, each electrode should have a controlling circuit, each
circuit should have at least two interface wires, all these elements
should be placed on a transparent plate with a matrix form. Figure 3 is a light
modulation device structure for hologram display suggested by NTT Docomo.
  \begin{figure}[h]%figure3
  \centering
  \includegraphics[width=3.5in, height=5in]{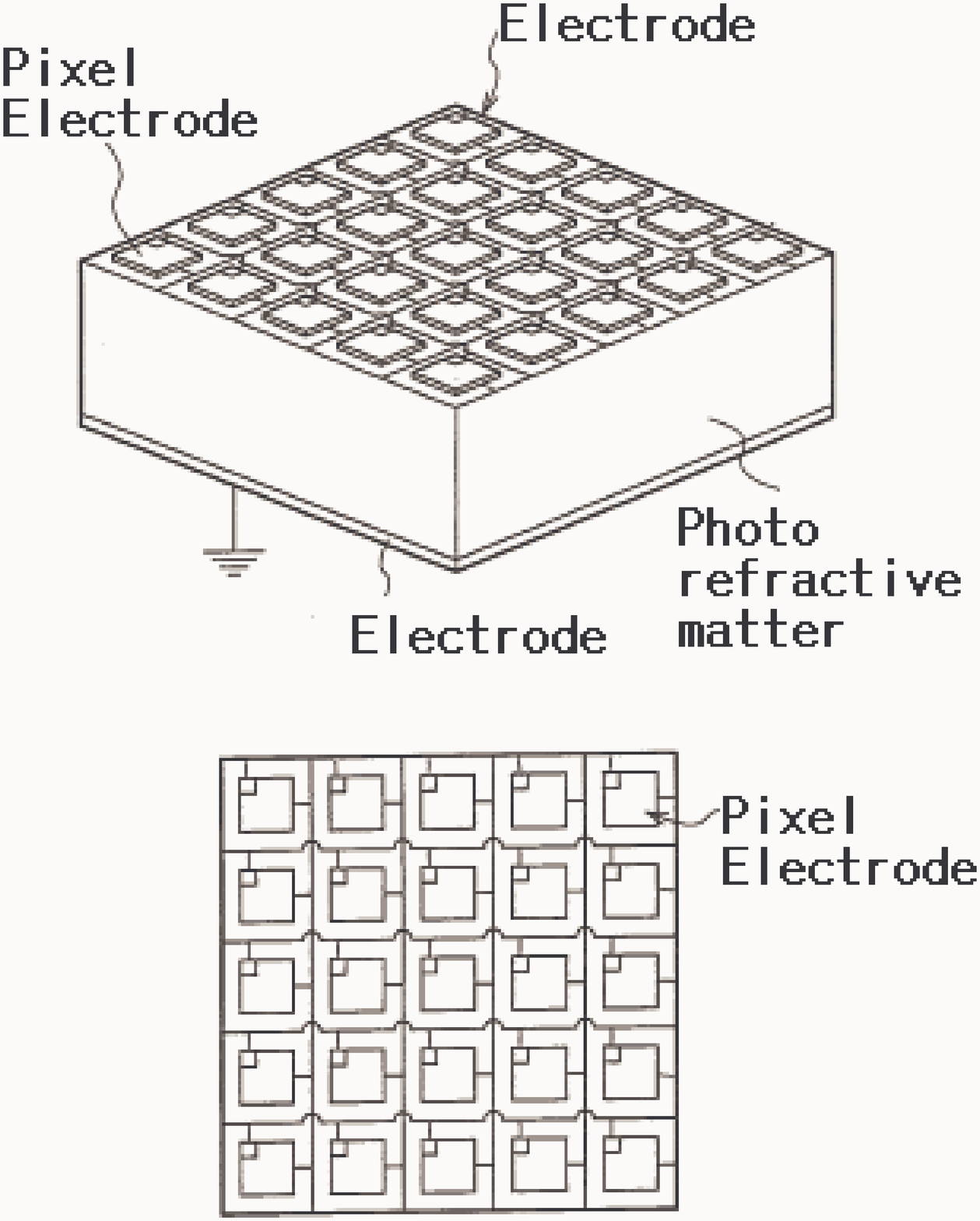}
  \caption{NTT Docomo's light modulator.}
  \end{figure}

To display a one-dimensional hologram, all the structures
mentioned above may not be placed on the displaying transparent
plate, except the pixel electrodes. Displaying the figure 4 clearly
doesn't  need the matrix of figure 3.

  \begin{figure}[h]%figure4
  \centering
  \includegraphics[angle=90, width=5.5in, height=3.5in]{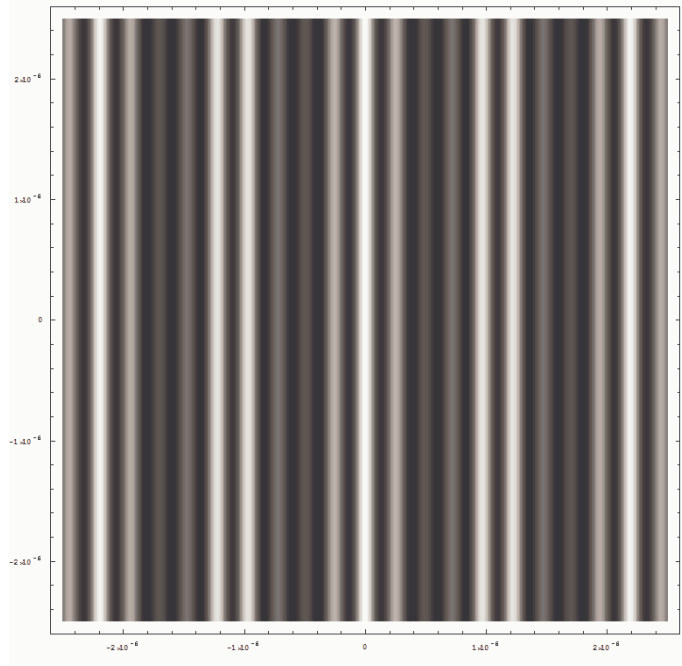}
  \caption{An example of one-dimensional Hologram.}
  \end{figure}

 Only the pixel electrodes are
needed to be placed for display and all other elements may be placed at the
edge of each electrode. This will improve the display resolution
almost to the limit of wiring technology. It seems that the recent
$60nm$ wiring technique is enough for the goal of $100nm$ resolution.

\subsection{The noise eliminating in hologram calculation}
A practical display device needs to consider about the problem of
image quality. The image quality is determined by resolution of image,
luminosity and noise.
The resolution problem shall not be discussed, because the holography is
intrinsically high resolution display, regardless of the modulator's
resolution.
And, the luminosity problems may be solved by multiple modulating of
phase modulation method. Then the noise remains.

The 4th term of Eq. (7) and term \{4\} of Eq. (12) are the noise
terms. These noises are caused from the assumption that
light modulation happens instantly at a surface. When
reproducing light is illuminated to hologram of Eq. (11), the energy
distribution $E$ of all modulated lights by hologram without
normalization is

\begin{equation}%equation 14
E=\left(
\begin{array}{ccccc}
1          & \alpha_2^2 & \alpha_3^2 & \alpha_4^2 & \cdots \\
\alpha_2^2 & (\alpha_2\alpha_2)^2 & (\alpha_2\alpha_3)^2 & (\alpha_2\alpha_4)^2 & \cdots \\
\alpha_3^2 & (\alpha_3\alpha_2)^2 & (\alpha_3\alpha_3)^2 & (\alpha_3\alpha_4)^2 & \cdots \\
\alpha_4^2 & (\alpha_4\alpha_2)^2 & (\alpha_4\alpha_3)^2 & (\alpha_4\alpha_4)^2 & \cdots \\
\cdots & \cdots & \cdots & \cdots & \cdots
\end{array}\right)
\end{equation}
When the number of image elements is $n$ and assuming that all image
elements have identical luminosity for convenience, the Eq. (14)
can be rewritten as $E=1+n\alpha^2+n\alpha^2+n^2\alpha^4$.
The first term is the 0th order term except diagonals, the second term
is the total luminosity of image, third term is the total luminosity
of conjugate image and fourth term is sum of 0th order diagonals and
noise term. when $\alpha\rightarrow 0$ with same $n$ the total luminosity of
image is sufficiently less than reference light, the energy of noise
term becomes negligible than the energy of image. This shows that the
noise term is especially important for the computer generated hologram
on a plain, and negligible when a volume hologram is used.
But, noise can be eliminated by simply throwing the noise term and using
row 1 and column 1 of Eq. (5), except diagonals. That is, instead of
the expression of Eq. (3)
\[I(\vec{r})= \left| \sum \alpha (\vec{r}) \exp{(i[f(\vec{r}) + \delta])} \right|^2 \]
, by adding row 1 and column 1 those elements are complex conjugates
with one another. So, it is expressed with cosine function.
\begin{equation} %equation 15
I(\vec{r})= 2\alpha_1(\vec{r})\sum\alpha(\vec{r})\cos(f_1(\vec{r})-f(\vec{r})+\delta_1-\delta)
\end{equation}
Applying  Eq. (15) to (11) to get expression of one-dimensional hologram,
\begin{equation} %equation 16
I(X)=
2\sum\alpha\cos(2\pi\frac{\textstyle(\hat{r}_1-\hat{r})\cdot\hat{x}X}{\lambda}+\delta_1-\delta)
\end{equation}
The negative value becomes possible, so, it needs different way for normalization.
This one-dimensional hologram can be called as multiplexed cosine diffracting lattice.

\subsection{Fast hologram computing}
One of the most big problem in hologram display device is its
tremendous data processing burden. When displaying 3-D image with a
hologram, there is no other way except improving the algorithm, but
when displaying 2-D image, it is possible to compute only partial
area of modulator, and can reuse its data on whole area to improve the
speed of computing. When using one-dimensional holography, this situation
becomes better. For two-dimensional hologram, all the hologram pixels
($\mbox{column H(ologram) pixels}\times\mbox{row H pixels}$)
must be computed by all the image pixels($\mbox{column I(mage)
  pixels}\times\mbox{row I pixels}$). But, For one-dimensional, just
one column of the hologram pixels($\mbox{column H pixels}$) are
computed by one column of the image pixels($\mbox{column I pixels}$),
and repeats this for number of the row line of image($\mbox{row I pixels}$).
 This increases computing speed by
\[
\frac{\mbox{column H pixels}\times\mbox{row H pixels}\times\mbox{column I
  pixels}\times\mbox{row I pixels}}
{\mbox{column H pixels}\times\mbox{column I pixels}\times\mbox{row I pixels}}
\]
equals
\[
\mbox{row H pixels}
\]
The aperture size of human eyes$^1$ are between $2mm$ to $8mm$. So, for
clean visuality, let the size of partial hologram to be $2mm$, and let
the hologram resolution to be $100nm$, then the one-dimensional
hologram may be computed 20,000 times faster than the 2-D displaying
two-dimensional hologram. But, current digital calculator may not be
able to handle the required computing burden for real time color
motion picture display.

Fortunately, there is other solution for one-dimensional hologram
 calculation. It is possible to make adjustable one-dimensional
 interference pattern then read it with photo-sensor array.

A coherent light started from a source is modulated and diffused by
light modulation device with input signal, this light is modulated
into multiple parallel rays by lens, and gain hologram output data by reading
interference patterns from those parallel rays with photo-sensor array.
This is shown on figure 5.
\begin{figure}[h]%figure5
  \centering
%  \leavemode
  \includegraphics[width=3.25in, height=4.5in]{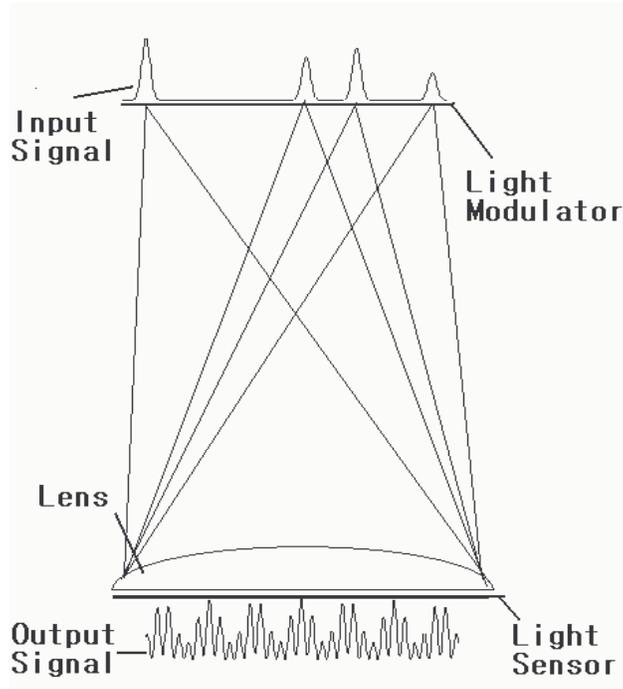}
  \caption{The optical computing structure.}
  \end{figure}
In this case, the calculation speed depends on the speed of
sufficient light gathering at the photo-sensor, a laser has
sufficient power with care of only generated heat.
The reference light was not indicated on figure 5. It is out of range
from radical axis, so, it can not be illuminated through the lens, it
should be illuminated diagonally from $z$ axis direction. In this case,
the noise removing method of
Eq. (16) can't be used, so, small values of $\alpha$ should be used. In
order to do so, A multi-layered one-dimensional hologram may be
used. The light modulation efficiency should be lowered at each
hologram, and the modulation is repeated with multiple layer.

 When looking at expression from Eq. (11),
\[ I(X)= \sum \alpha \exp{( -2 \pi
  i[\frac{\hat{r}\cdot\hat{x}}{\lambda}X +\delta ])}
\times \sum \alpha \exp{( 2 \pi  i
[\frac{\hat{r}\cdot\hat{x}}{\lambda}X +\delta ])}\]
it shows that only $\hat{r}\cdot\hat{x}$ values are used for
one-dimensional hologram calculation.
Therefore, the structure of figure 5 can be applied to all
latitude lines regardless of $\hat{r}\cdot\hat{y}$.
Also because, reducing $\hat{r}\cdot\hat{x}$ value and properly increasing
$X$ value results in same, so, input pixels can be changed to more
paraxial, and at the same time, it makes the size of the
photosensor array larger.
The method in figure 5 can be formed in a thin shape with tens
thousand pixel lineal CCD in one-dimensional holography, but when applied to
two-dimensional hologram, it would encounter the problems of embodying
in a thick shape, illuminating the reference light very out of ranged
from radical axis, and making hundreds million pixel CCD.

\subsection{The comparison with diffraction lattice}

As mentioned above, a one-dimensional hologram can be regarded as a
multiplexed diffraction lattice, too. So, the comparison of one-dimensional
and diffraction lattice is considerable.

When examining the calculation speed of diffraction lattice to display
an image, for a diffraction lattice, each column of the lattice line
pixel ( $\mbox{column L pixels}$) should be computed by each pixels of the
image, and this should be repeated for the number of the row lines of
image ( $\mbox{row I pixels}$), then repeated again for the number of
the column lines of image ( $\mbox{column I pixels}$). This amount of
calculation equals to that of the one-dimensional hologram.

Considering the computing speed,
diffraction lattice is not bad, but there are other problems in
diffraction lattice method. The light modulator for displaying the
diffraction lattice should be changed for each image pixels. When one
line image of one-dimensional hologram consists of 2000 pixels,
the modulator for diffraction lattice should be reconfigured 2000
times more than one-dimensional hologram. This means that the
light modulator and all the elements of figure 5 should have 2000
times faster speed than those of one-dimensional hologram.

To compare the data transfer rate, let us assume that light
modulator consists of $1mm\times 1mm$ parts 
by $100nm$ resolution, the displaying image consists
of $2000\times 2000$ pixels, and let us choose the frame rate of 48 frames
per a second(24 is traditional frame rate, but a hologram or a lattice
can express only one color at a time, thus some extra frames are
needed. It seems that 72 monochrome frames are not required for 24
color frames, when 6 frames are used for three times of shape refreshing, and two times of
color refreshing, 48 frames are enough. 48 is chosen because it is
about 50 that is easy to handle.), then the time limit for a frame of
two-dimensional hologram is about $1s/50=20ms$, for a line frame of one-dimensional
hologram is about $1s/50/2000=10\mu s$, and for a dot frame of the
diffraction lattice is $1s/50/2000/2000=5ns$.(A $2000\times 2000$
pixel image is quite high resolution for plain pictures, but it is only not so
bad resolution for eyeglasses type display devices of wide visual angle.) 
And, with $1mm/100nm = 10000$, the transfer rates are calculated as
$10000\times 10000/20ms = 5$ giga times per a second for two-dimensional
hologram, $10000/10\mu s = 1$ giga times for one-dimensional hologram, and
$10000/5ns = 2$ tera times for diffraction lattice. These results show
that the one-dimensional holography is most efficient.

Some other ways of using diffraction lattice exist they avoid calculation
and transmission of data. A material which self
arranges its fringe by voltage has been known, and the method of
using acoustic wave as the diffraction lattice also has been known. But,
the arrangement speeds of these methods seem difficult
to meet $5ns$ of time limit,
because the state of molecules in that kind of material should be
determined by their neighbour molecules, and the informations are exchanged
with the speed of acoustic wave. The acoustic wave method may be
considerable for one-dimensional holography, if it is capable of
expressing $100nm$ resolution.
 
\section{Conclusion}

The one-dimensional holography is a new display method which has
balanced characteristics between conventional two-dimensional holography
and diffraction lattice method. This is a theoretical method yet, and thus it
seems that there are no precedent and few references.
But, this is not a unique theory, this is an application of common theory
for a special problem, thus, this could be theoretically verified easily.

Many researcher's dedications are required
for practical use of one-dimensional holography. Especially,
the research for a fast responsive light modulating material seems
essential. As modulating method, phase modulating or polarization
modulating material may be adequate. Also, more precise design of
optical calculator is required. Other computing methods like analog
computing device or faster DSP could be researched. And, many
others may also be needed.

%% Code for appendices and equation numbers
%\appendix

%\section*{Appendix A: Sample}
%\setcounter{equation}{0}
%\renewcommand{\theequation}{A{\arabic{equation}}}

%\begin{equation} a+b=c.
%\end{equation}

%\section*{Appendix B: Sample}
%\setcounter{equation}{0}
%\renewcommand{\theequation}{B{\arabic{equation}}} %change B as needed
%\begin{equation}
%x-y=z.
%\end{equation}

\end{document}